\documentclass[
preprint,
showpacs
]{revtex4-1}
\usepackage{amsthm}
\theoremstyle{plain}
\newtheorem{theorem}{Theorem}%[section]
\newtheorem*{theorem*}{Theorem}

\newtheorem*{definition*}{Definition}

\newtheorem*{lemma*}{Lemma}

\usepackage{braket}
\usepackage{delarray}
\usepackage{here}

\usepackage{amsmath}
\usepackage{txfonts}

\usepackage[dvipdfmx]{graphicx}
\usepackage{bm}
\usepackage{color}
\usepackage{ulem}

\usepackage{physics}

\newcommand{\ba}{\begin{array}}
\newcommand{\ea}{\end{array}}
\newcommand{\bmat}{\left(\begin{array}}
\newcommand{\emat}{\end{array}\right)}
\newcommand{\no}{\nonumber}

\newcommand{\be}{\begin{eqnarray}}
\newcommand{\ee}{\end{eqnarray}}

\begin{document}
\title{Replica bound for Ising spin glass models in one dimension}
\author{Manaka Okuyama$^1$\footnote{corresponding author: manaka.okuyama.d2@tohoku.ac.jp}}
\author{Masayuki Ohzeki$^{1,2,3}$}
\affiliation{$^1$Graduate School of Information Sciences, Tohoku University, Sendai 980-8579, Japan}
\affiliation{$^2$Department of Physics, Institute of Science Tokyo, Tokyo 152-8551, Japan}
\affiliation{$^3$Sigma-i Co., Ltd., Tokyo 108-0075, Japan} %\\

\begin{abstract}  
The interpolation method is a powerful tool for rigorous analysis of mean-field spin glass models, both with and without dilution.
In this study, we show that the interpolation method can be applied to Ising spin glass models in one dimension, such as a one-dimensional chain and a two-leg ladder.
In one dimension, the replica symmetric (RS) cavity method is naturally expected to be rigorous for Ising spin glass models.
Using the interpolation method, we rigorously prove that the RS cavity method provides lower bounds on the quenched free energies of Ising spin glass models in one dimension at any finite temperature in the thermodynamic limit.

\end{abstract}
\date{\today}
\maketitle

%%%%%%%%%%%%%%%%%%%%%%%%%%%%%%%%%%%%%%%%%%%%%%%%%%%%%%%%%%%%%%%%%%%%%%%%%%%
%%%%%%%%%%%%%%%%%%%%%%%%%%%%%%%%%%%%%%%%%%%%%%%%%%%%%%%%%%%%%%%%%%%%%%%%%%%
\section{Introduction}

Spin-glass models describe magnetic materials with random interactions and have been studied extensively from both statistical mechanics and information science perspectives~\cite{Nishimori,MM}.
The standard analytical method for spin glass models is the replica method~\cite{MM} and its application to mean-field spin glass models has achieved significant success.
The concept of replica symmetry breaking~\cite{Parisi}, derived from the replica method, has significantly deepened our understanding of randomness.
It is important to note that, while the results obtained using the replica method are widely regarded as correct in many cases, the replica method itself is not mathematically rigorous.

On the other hand, mathematically rigorous analysis of mean-field spin glass models has recently made remarkable progress through the interpolation method~\cite{GT}.
This method introduces a parameter interpolating between the random interaction and the auxiliary random field. 
By examining the response of the system to the interpolating parameter, the free energy of mean-field spin glass models can be evaluated by a quantity calculated from the auxiliary random field.
This procedure significantly simplifies the analysis of mean-field spin glass models.
As a result, it is possible to obtain an inequality for the free energy (it often coincides with the result of the replica method).
Based on the interpolation method, the full-step replica symmetry breaking solution~\cite{Parisi} has been proven to be an exact solution for the free energy of the SK model~\cite{Guerra,Talagrand,Panchenko}.
The interpolation method is also a powerful tool for diluted mean-field spin glass models, demonstrating that the replica symmetric (RS) and one-step replica symmetry breaking cavity methods~\cite{MG} provide lower bounds on the free energies of these models~\cite{FL,FLT,PT}.
Efforts to obtain inverse bounds for diluted mean-field spin glass models are ongoing~\cite{GT2,Panchenko2,Panchenko3,Panchenko4,OP,DSS,LO,Harangi}.

Unfortunately, the interpolation method does not work well in finite dimensions.
In one dimension, the Ising spin glass models have been analyzed using the replicated transfer matrix method~\cite{DVP,LFN,WM,BM,LMR}.
Notably, a recent study~\cite{LMR} has proved that the RS cavity method is rigorous for the Ising spin glass model on a one-dimensional chain with free boundary condition (by computing the $n$-replicated partition function for any real number $n$ and then rigorously taking the limit $n\to0$).
However, to the best of our knowledge, the interpolation method has not been applied in one dimension.

In the present study, we used the interpolation method~\cite{GT} to analyze the Ising spin glass models in one dimension, such as a one-dimensional chain and a two-leg ladder. 
The key element is the use of dilution variables introduced in the analysis of diluted mean-field models~\cite{FL} as the interpolating parameter. 
However, our interpolation method employs a slightly different parameter setup compared to that of a previous study.
In diluted mean field models~\cite{FL}, dilution variables following Poisson distributions are introduced separately for the random interaction and the auxiliary field.
 In contrast, in the present study, a common dilution variable following a Bernoulli distribution is introduced for a set of the random interaction and the auxiliary random field.
While this difference might seem insignificant, it plays a crucial role in one-dimensional Ising spin glass models.
Then, we rigorously prove that the RS cavity method provides lower bounds on the free energies of the Ising spin glass models in one dimension.
The reason why this proof works well for one-dimensional systems is that our interpolation method is closely related to Oguchi's approximate free energy~\cite{Oguchi} and is particularly suited to such systems (details are provided in Sec. IV). 
In contrast, the interpolation method for diluted mean field models~\cite{FL} is incompatible with one-dimensional systems and does not perform well in the present case.

The rest of this paper is organized as follows.
In Sec. II, we define the Ising spin glass models in one dimension and state the main results.
In Sec. III, we use the interpolation method to prove that the RS cavity method provides lower bounds on the free energies of these models.
Our results are discussed in Sec. IV.
For completeness, the derivation of the RS cavity formula is provided in Appendices A and B.

%%%%%%%%%%%%%%%%%%%%%%%%%%%%%%%%%%%%%%%%%%%%%%%%%%%%%%%%%%%%%%%%%
\section{Definitions and results}
\subsection{One-dimensional chain}
Let us consider the Ising spin glass model on a one-dimensional chain with periodic boundary conditions:
\be
H_{N}^\mathrm{chain}&=&- \sum_{i=1}^{N} J_{i} \sigma_i \sigma_{i+1}  - \sum_{i=1}^{N}  H_i\sigma_i ,
\ee
where $\sigma_i=\pm1$, $J_i$ are i.i.d. random variables following any symmetric distribution $P(J_i)=P(-J_i)$, and $H_i$ are i.i.d. random variables following any distribution (the probability distribution of $H_i$ is allowed to present no randomness, for example, a uniform magnetic field).
We assume that these distributions are regular enough to guarantee that all expressions below are well-defined.

The partition function is defined as
\be
Z_N^\mathrm{chain}&=&\Tr(e^{-\beta H_{N}^\mathrm{chain}}) ,
\ee
where $\Tr(\cdots)$ denotes the summation over all spin variables, and $\beta$ is the inverse temperature.
The thermal average $\langle \cdots \rangle$ is given by
\be
\langle g(\sigma) \rangle&=&\frac{\Tr\qty( g(\sigma) e^{-\beta H_{N+1}^\mathrm{chain} })}{Z_N^\mathrm{chain}} .
\ee
The quenched free energy is
\be
f_N^\mathrm{chain}&=&-\frac{1}{\beta N}\mathbb{E}\left[\log Z_N^\mathrm{chain} \right],
\ee
where $\mathbb{E}\left[\cdots\right]$ denotes the expectation with respect to all random variables.
The thermodynamic limit of the quenched free energy is defined as
\be
f^\mathrm{chain}&=&\lim_{N\to\infty}f_N^\mathrm{chain}.
\ee
The existence of the thermodynamic limit of the quenched free energy is proved through the conventional approach~\cite{Vuillermat,PF}.

By applying the RS cavity method~\cite{MM} to the Ising spin glass model on a one-dimensional chain, we obtain the RS cavity formula~\cite{Matsubara,KF} (see Appendix A for the derivation)
\be
f_\mathrm{RS}^\mathrm{chain}&=&-\frac{1}{\beta}\mathbb{E}\qty[\log \frac{2\cosh\beta J_{} \cosh\beta (\hat{h}+H)}{\cosh \beta\hat{h}_{}}], \label{chain-cavity-formula}
\ee 
where $J$ is a random variable following the same distribution as $J_i$, $H$ is a random variable following the same distribution as $H_i$, and the cavity field $\hat{h}$ satisfies
\be
\tanh\beta\hat{h}&\stackrel{d}{=}&\tanh \beta J \tanh\beta(H+\hat{h}_{}') ,\label{chain-cavity-eq}
\ee
where the equality holds in distribution and $\hat{h}_{}'$ is a random variable following the same distribution as $\hat{h}$.

Then, our first main result is as follows:
%%%%%%%%%%%%%%%%%%%%%%%%%%%%%%%%%%%%%%%%%%%%%%%%%%%%%%%%%%%%%%%
%%%%%%%%%%%%%%%%%%%%%%%%%%%%%%%%%%%%%%%%%%%%%%%%%%%%%%%%%%%%%%%
\begin{theorem}\label{theorem1}
In the thermodynamic limit, the quenched free energy of the Ising spin glass model on a one-dimensional chain is bounded below by the RS cavity formula
\be
f_\mathrm{RS}^\mathrm{chain}\le  f^\mathrm{chain} .
\ee
\end{theorem}
%%%%%%%%%%%%%%%%%%%%%%%%%%%%%%%%%%%%%%%%%%%%%%%%%%%%%%%%%%%%%%%
%%%%%%%%%%%%%%%%%%%%%%%%%%%%%%%%%%%%%%%%%%%%%%%%%%%%%%%%%%%%%%%

The proof of Theorem \ref{theorem1} is based on the interpolation method in Ref.~\cite{FL}, where it was proven that the RS cavity method gives lower bounds on the free energies for dilute mean-field models.
We introduce Bernoulli-type dilution variables that interpolate between the one-dimensional spin glass model and the cavity field model.
It is worth noting that our interpolation method differs slightly from the one used in the previous study~\cite{FL} in that we use a common dilution variable for a set of the random interaction and the cavity field.
Using this interpolating parameter, we define the interpolating free energy.
By investigating the derivative of the interpolating free energy with respect to the interpolating parameter, we find that a quantity derived from the interpolating free energy coincides with the result of the RS cavity method (see Eq. (\ref{chain-info-t=0})).
This key relation implies that, to prove Theorem \ref{theorem1}, it is sufficient to show that the second term in Eq. (\ref{identity}) is positive.
This can be demonstrated using the exponential decay of correlation functions in one dimension.
Consequently, we establish that the RS cavity method gives a lower bound on the free energy of the Ising spin glass model on a one-dimensional chain.

%%%%%%%%%%%%%%%%%%%%%%%%%%%%%%%%%%%%%%%%%%%%%%%%%%%%%%%%%%%%%%%
%%%%%%%%%%%%%%%%%%%%%%%%%%%%%%%%%%%%%%%%%%%%%%%%%%%%%%%%%%%%%%%
\subsection{Two-leg ladder}

Next, we consider the Ising spin glass model on a two-leg ladder with periodic boundary conditions:
\be
H_N^\mathrm{ladder}&=&-\sum_{i=1}^N L_{i}\sigma_{i,1}\sigma_{i,2} - \sum_{i=1}^N(K_{i,1}\sigma_{i,1}\sigma_{i+1,1} + K_{i,2}\sigma_{i,2}\sigma_{i+1,2}),
\ee
where $\sigma_{i,1}=\pm1$, $\sigma_{i,2}=\pm1$, $K_{i,1}$ are i.i.d. random variables following any symmetric distribution $P(K_{i,1})=P(-K_{i,1})$, $K_{i,2}$ are i.i.d. random variables following any symmetric distribution $P(K_{i,2})=P(-K_{i,2})$, and $L_i$ are i.i.d. random variables following any distribution.
The probability distribution of $L_i$ is allowed to present no randomness.

The partition function is given by 
\be
Z_N^\mathrm{ladder}&=&\Tr\qty(e^{-\beta H_N^\mathrm{ladder}}).
\ee
The quenched free energy and its thermodynamic limit are defined as
\be
f_N^\mathrm{ladder}&=&-\frac{1}{\beta N}\mathbb{E}\left[\log Z_N^\mathrm{ladder} \right],
\\
f^\mathrm{ladder}&=&\lim_{N\to\infty}f_N^\mathrm{ladder}.
\ee

By applying the RS cavity method to the Ising spin glass model on a two-leg ladder, we obtain the RS cavity formula (see Appendix B for the derivation)
\be
f_\mathrm{RS}^\mathrm{ladder}&=&-\frac{1}{\beta}\mathbb{E}\qty[\log \frac{4\cosh\beta K_1 \cosh\beta K_2\cosh\beta(L+\hat{l})}{ \cosh\beta\hat{l} }],\label{ladder-cavity-formula}
\ee
where $K_1$ is a random variable following the same distribution as $K_{i,1}$, $K_2$ is a random variable following the same distribution as $K_{i,2}$, $L$ is a random variable following the same distribution as $L_i$, and the cavity coupling $\hat{l}$ satisfies
\be
\tanh\beta \hat{l}  &\stackrel{d}{=}& \tanh\beta K_1 \tanh\beta K_2 \tanh\beta (L+\hat{l}') , \label{ladder-cavity-eq}
\ee
where the equality holds in distribution, and $\hat{l}_{}'$ is a random variable following the same distribution as $\hat{l}$.

The second main result is as follows:
%%%%%%%%%%%%%%%%%%%%%%%%%%%%%%%%%%%%%%%%%%%%%%%%%%%%%%%%%%%%%%%
\begin{theorem}\label{theorem2}
The thermodynamic limit of the quenched free energy of the Ising spin glass model on the two-leg ladder is bounded by the RS cavity formula
\be
f_\mathrm{RS}^\mathrm{ladder}\le  f^\mathrm{ladder} .
\ee
\end{theorem}

To obtain Theorem \ref{theorem2}, we introduce Bernoulli-type dilution variables that interpolate between the Ising spin glass model on the two-leg ladder and the cavity coupling model.
Then, the proof of Theorem \ref{theorem2} is almost same as that of Theorem \ref{theorem1}.

%%%%%%%%%%%%%%%%%%%%%%%%%%%%%%%%%%%%%%%%%%%%%%%%%%%%%%%%%%%%%%%
\section{Proof}
%%%%%%%%%%%%%%%%%%%%%%%%%%%%%%%%%%%%%%%%%%%%%%%%%%%%%%%%%%%%%%%
\subsection{Proof of Theorem 1}
We define the interpolating Hamiltonian as 
\be
H_{N}^\mathrm{chain}(t)&=&- \sum_{i=1}^{N} \qty(J_i M_{i}(t)  \sigma_i \sigma_{i+1} + (1-M_{i}(t))(\hat{h}_{i,1}\sigma_i+\hat{h}_{i,2}\sigma_{i+1})  ) -\sum_{i=1}^{N} H_i \sigma_i,
\\
H_{N}^\mathrm{chain}(1)&=&H_{N}^\mathrm{chain},
\\
H_{N}^\mathrm{chain}(0)&=&- \sum_{i=1}^N(H_i+\hat{h}_{i,1}+\hat{h}_{i-1,2})\sigma_i ,
\ee
where $M_{i}(t)$ are i.i.d. Bernoulli variables with average $t$ $(0\le t\le1)$, and  $\hat{h}_{i,1}$ and $\hat{h}_{i,2}$ are random variables following the same distribution as the cavity field $\hat{h}$ (see Eq. (\ref{chain-cavity-eq})).
We define the interpolating quenched free energy as
\be
f_N^\mathrm{chain}(t)&=&-\frac{1}{\beta N}\mathbb{E}\left[\log Z_N^\mathrm{chain}(t) \right].  \label{f(t)}
\ee
At $t=1$, we recover the free energy of the Ising spin glass model on the one-dimensional chain
\be
f_N^\mathrm{chain}(1)&=&f_N^\mathrm{chain}.
\ee

For any function $g(x)$, the following identity holds:
\be
\dv{}{t}\mathbb{E}\qty[g\qty(M_i(t))]&=&g(1)-g(0).
\ee
Using this identity, we have
\be
\dv{}{t}f_N^\mathrm{chain}(t)&=&-\frac{1}{\beta N}\sum_{i=1}^N\mathbb{E}\qty[  \log \frac{\langle e^{\beta J_{i} \sigma_i \sigma_{i+1}}\rangle_{H_{\backslash i}^\mathrm{chain}(t)} }{ \langle e^{\beta (\hat{h}_{i,1}\sigma_i+\hat{h}_{i,2}\sigma_{i+1})}\rangle_{H_{\backslash i}^\mathrm{chain}(t)} }  ]
\no\\
&=&-\frac{1}{\beta }\mathbb{E}\qty[  \log \frac{\langle e^{\beta J_{i} \sigma_i \sigma_{i+1}}\rangle_{H_{\backslash i}^\mathrm{chain}(t)} }{ \langle e^{\beta (\hat{h}_{i,1}\sigma_i+\hat{h}_{i,2}\sigma_{i+1})}\rangle_{H_{\backslash i}^\mathrm{chain}(t)} }  ],
\ee
where we have applied the translation invariance property, and $\langle\cdots \rangle_{H_{\backslash i}^\mathrm{chain}(t)} $ denotes the thermal average with respect to $H_{\backslash i}^\mathrm{chain}(t)$.
Here, $H_{\backslash i}^\mathrm{chain}(t)$ is defined as
\be
H_{\backslash i}^\mathrm{chain}(t)&=& H_{N}^\mathrm{chain}(t) +J_i M_{i}(t)  \sigma_i \sigma_{i+1} + (1-M_{i}(t))(\hat{h}_{i,1}\sigma_i+\hat{h}_{i,2}\sigma_{i+1}). \label{chain-inter-H-free}
\ee
Note that  $H_{\backslash i}^\mathrm{chain}(t)$ does not contain $J_i$, $\hat{h}_{i,1}$, and $\hat{h}_{i,2}$. 
The lattice structure of $H_{\backslash i}^\mathrm{chain}(t)$ is that of a one-dimensional chain with free boundary conditions instead of periodic boundary conditions.

By using the identity $\exp(\beta J_i \sigma_i\sigma_{i+1})=\cosh\beta J_i (1+\sigma_i\sigma_{i+1}\tanh\beta J_i)$ and Eq. (\ref{chain-cavity-eq}), we obtain
\be
f_N^\mathrm{chain}(0)+\dv{}{t}f_N^\mathrm{chain}(t)\big|_{t=0}&=&f_\mathrm{RS}^\mathrm{chain}. \label{chain-info-t=0}
\ee
This equation implies that the information at $t=0$ coincides with the result of the RS cavity method.
This structure is also common in the analysis of mean-field models using the interpolation method~\cite{Guerra,FL}.
By using the fundamental theorem of calculus and Eq. (\ref{chain-info-t=0}), we have
\be
f_N^\mathrm{chain}(1)&=&f_N^\mathrm{chain}(0)+\int_0^1 dt\dv{}{t}f_N^\mathrm{chain}(t)
\no\\
&=&f_\mathrm{RS}^\mathrm{chain}+\int_0^1 dt\qty(\dv{}{t}f_N^\mathrm{chain}(t) -\dv{}{t}f_N^\mathrm{chain}(t)\big|_{t=0}) . \label{identity}
\ee
Thus, to obtain Theorem 1, it is sufficient to prove 
\be
\int_0^1 dt\qty(\dv{}{t}f_N^\mathrm{chain}(t) -\dv{}{t}f_N^\mathrm{chain}(t)\big|_{t=0}) \ge \mathcal{O}(N^{-1}).  \label{target1}
\ee

We obtain
\be
&&-\beta\qty (\dv{}{t}f_N^\mathrm{chain}(t) -\dv{}{t}f_N^\mathrm{chain}(t)\big|_{t=0})
\no\\
&=&  \mathbb{E}\qty[  \log (1+\tanh \beta J_i \langle \sigma_i \sigma_{i+1} \rangle_{H_{\backslash i}^\mathrm{chain}(t)} ) ]
- \mathbb{E}\qty[   \log \langle  (1+\tanh\beta \hat{h}_{i,1}  \sigma_i   ) (1+\tanh\beta \hat{h}_{i,2}  \sigma_{i+1}   ) \rangle_{H_{\backslash i}^\mathrm{chain}(t)}  ]
\no\\
&&+\mathbb{E}\qty[  \log (1+\tanh \beta J_i \tanh\beta(H_i+\hat{h}_{i,1}) \tanh\beta(H_{i+1}+\hat{h}_{i+1,1}) ) ] .\label{first-t-0}
\ee
Here, we recall that, in the thermal average $\langle \cdots \rangle_{H_{\backslash i}^\mathrm{chain}(t)}$, sites $i$ and $i+1$ are both ends of a one-dimensional chain with free boundary conditions.
This fact implies that sites $i$ and $i+1$ are connected with probability $t^{N-1}$, given that the existence probability of each interaction is $t$.
In other words, the correlation function is decoupled,
\be
\langle \sigma_i\sigma_{i+1} \rangle_{H_{\backslash i}^\mathrm{chain}(t)}=\langle \sigma_i \rangle_{H_{\backslash i}^\mathrm{chain}(t)}\langle \sigma_{i+1} \rangle_{H_{\backslash i}^\mathrm{chain}(t)},
\ee
with probability $1-t^{N-1}$.
Then, we have
\be
&&- \mathbb{E}\qty[   \log \langle  (1+\tanh\beta \hat{h}_{i,1}  \sigma_i   ) (1+\tanh\beta \hat{h}_{i,2}  \sigma_{i+1}   ) \rangle_{H_{\backslash i}^\mathrm{chain}(t)}  ]
\no\\
&=&- \mathbb{E}\qty[   \log   (1+\tanh\beta \hat{h}_{i,1}  \langle \sigma_i \rangle_{H_{\backslash i}^\mathrm{chain}(t)}  ) \qty(1+\tanh\beta \hat{h}_{i,2} \langle \sigma_{i+1} \rangle_{H_{\backslash i}^\mathrm{chain}(t)}  )   ] +\mathcal{O}(t^{N-1})
\no\\
&=&- 2\mathbb{E}\qty[   \log   (1+\tanh \beta J_i \tanh\beta(H_i+\hat{h}_{i,1})   \langle \sigma_i \rangle_{H_{\backslash i}^\mathrm{chain}(t)}  )  ] +\mathcal{O}(t^{N-1}) , \label{dilute-evalu}
\ee
where we have used Eq. (\ref{chain-cavity-eq}) in the last equality.
By substituting Eq. (\ref{dilute-evalu}) into Eq. (\ref{first-t-0}), we obtain
\be
&&-\beta\qty (\dv{}{t}f_N^\mathrm{chain}(t) -\dv{}{t}f_N^\mathrm{chain}(t)\big|_{t=0})
\no\\
&=&  \mathbb{E}\qty[  \log (1+\tanh \beta J_i \langle \sigma_i \sigma_{i+1} \rangle_{H_{\backslash i}^\mathrm{chain}(t)} ) ]
- 2\mathbb{E}\qty[   \log   (1+\tanh \beta J_i \tanh\beta(H_i+\hat{h}_{i,1})   \langle \sigma_i \rangle_{H_{\backslash i}^\mathrm{chain}(t)}  )  ]
\no\\
&&+\mathbb{E}\qty[  \log (1+\tanh \beta J_i \tanh\beta(H_i+\hat{h}_{i,1}) \tanh\beta(H_{i+1}+\hat{h}_{i+1,1}) ) ]  +\mathcal{O}(t^{N-1}) .
\ee
Given that
\be
|\tanh \beta J_i \langle \sigma_i \sigma_{i+1} \rangle_{H_{\backslash i}^\mathrm{chain}(t)}|&<&1,
\\
|\tanh \beta J_i \tanh\beta(H_i+\hat{h}_{i,1})\langle \sigma_i \rangle_{H_{\backslash i}^\mathrm{chain}(t)}|&<&1 ,
\\
|\tanh \beta J_i \tanh\beta(H_i+\hat{h}_{i,1}) \tanh\beta(H_{i+1}+\hat{h}_{i+1,1}) |&<&1,
\ee
we can expand the logarithm of the three terms as
\be
&&-\beta\qty (\dv{}{t}f_N^\mathrm{chain}(t) -\dv{}{t}f_N^\mathrm{chain}(t)\big|_{t=0})
\no\\
&=& \sum_{n=1}^\infty \frac{(-1)^{n-1}}{n}\mathbb{E}\left[  \qty(\tanh \beta J_i \langle \sigma_i \sigma_{i+1} \rangle_{H_{\backslash i}^\mathrm{chain}(t)} )^{n} -2\qty(\tanh \beta J_i \tanh\beta(H_i+\hat{h}_{i,1})   \langle \sigma_i \rangle_{H_{\backslash i}^\mathrm{chain}(t)})^n \right.
\no\\
&&\left. + \qty(\tanh \beta J_i \tanh\beta(H_i+\hat{h}_{i,1}) \tanh\beta(H_{i+1}+\hat{h}_{i+1,1}) )^n \right]+\mathcal{O}(t^{N-1}) ,
\no\\
&=&-\sum_{n=1}^\infty\frac{1}{2n}\mathbb{E}\qty[\tanh^{2n} \beta J_i ]
\no\\
&&\qty(\mathbb{E}\qty[  \langle \sigma_i \sigma_{i+1} \rangle_{H_{\backslash i}^\mathrm{chain}(t)}^{2n}  ]-2\mathbb{E}\qty[  \tanh^{2n}\beta(H_i+\hat{h}_{i,1})]\mathbb{E}[   \langle \sigma_i \rangle_{H_{\backslash i}^\mathrm{chain}(t)}^{2n}  ] +\mathbb{E}\qty[  \tanh^{2n}\beta(H_i+\hat{h}_{i,1})]^2)
\no\\
&&+\mathcal{O}(t^{N-1}) , \label{first-t-0-2}
\ee
where we have used $\mathbb{E}[\tanh^{2n-1} \beta J_i ]=0$ for any positive integer $n$ that follows from $P(J_i)=P(-J_i)$.
Again, we use the fact that sites $i$ and $i+1$ are connected with probability $t^{N-1}$, which implies
\be
\mathbb{E}\qty[  \langle \sigma_i \sigma_{i+1} \rangle_{H_{\backslash i}^\mathrm{chain}(t)}^{2n}  ]&=&\mathbb{E}\qty[  \langle \sigma_i  \rangle_{H_{\backslash i}^\mathrm{chain}(t)}^{2n}  ]^2+\mathcal{O}(t^{N-1}). \label{dilute-evalu-2}
\ee
By applying Eq. (\ref{dilute-evalu-2}) to Eq. (\ref{first-t-0-2}), we obtain 
\be
&&-\beta\qty (\dv{}{t}f_N^\mathrm{chain}(t) -\dv{}{t}f_N^\mathrm{chain}(t)\big|_{t=0})
\no\\
&=&-\sum_{n=1}^\infty\frac{1}{2n}\mathbb{E}\qty[\tanh^{2n} \beta J_i ] \mathbb{E}\qty[  \langle \sigma_i  \rangle_{H_{\backslash i}^\mathrm{chain}(t)}^{2n}  -  \tanh^{2n}\beta(H_i+\hat{h}_{i,1})]^2
+\mathcal{O}(t^{N-1}) 
\no\\
&\le&\mathcal{O}(t^{N-1}) . 
\ee
Thus, we have
\be
f_N^\mathrm{chain}=f_N^\mathrm{chain}(1)
&=&f_\mathrm{RS}^\mathrm{chain}+\int_0^1 dt\qty(\dv{}{t}f_N^\mathrm{chain}(t) -\dv{}{t}f_N^\mathrm{chain}(t)\big|_{t=0}) 
\no\\
&\ge&f_\mathrm{RS}^\mathrm{chain}+\int_0^1 dt \mathcal{O}(t^{N-1}) 
\no\\
&=&f_\mathrm{RS}^\mathrm{chain}+ \mathcal{O}(N^{-1}) .
\ee
Finally, taking the thermodynamic limit, we obtain
\be
  f^\mathrm{chain} \ge f_\mathrm{RS}^\mathrm{chain} ,
\ee
which proves Theorem 1.

%%%%%%%%%%%%%%%%%%%%%%%%%%%%%%%%%%%%%%%%%%%%%%%%%%%%%%%%%%%%%%%
%%%%%%%%%%%%%%%%%%%%%%%%%%%%%%%%%%%%%%%%%%%%%%%%%%%%%%%%%%%%%%%
\subsection{Proof of Theorem 2}
We define the interpolating Hamiltonian as 
\be
H_N^\mathrm{ladder}(t)&=&-\sum_{i=1}^N L_{i}\sigma_{i,1}\sigma_{i,2} - \sum_{i=1}^N M_i(t)(K_{i,1}\sigma_{i,1}\sigma_{i+1,1} + K_{i,2}\sigma_{i,2}\sigma_{i+1,2}),
\no\\
&&-\sum_{i=1}^N(1-M_i(t))(\hat{l}_{i,1}\sigma_{i,1}\sigma_{i,2}  + \hat{l}_{i,2}\sigma_{i+1,1}\sigma_{i+1,2}  ),
\\
H_N^\mathrm{ladder}(1)&=&H_N^\mathrm{ladder},
\\
H_N^\mathrm{ladder}(0)&=&-\sum_{i=1}^N (L_{i}+\hat{l}_{i,1}+\hat{l}_{i-1,2}) \sigma_{i,1}\sigma_{i,2},
\ee
where $M_{i}(t)$ are i.i.d. Bernoulli variables with average $t$ $(0\le t\le1)$, and  $\hat{l}_{i,1}$ and $\hat{l}_{i,2}$ are random variables following the same distribution as the cavity coupling $\hat{l}$ (see Eq. (\ref{ladder-cavity-eq})).
We define the interpolating quenched free energy as
\be
f_N^\mathrm{ladder}(t)&=&-\frac{1}{\beta N}\mathbb{E}\left[\log Z_N^\mathrm{ladder}(t) \right].
\ee
Note that
\be
f_N^\mathrm{ladder}(1)&=&f_N^\mathrm{ladder}.
\ee
Then, we have
\be
\dv{}{t}f_N^\mathrm{ladder}(t)
&=&-\frac{1}{\beta N}\sum_{i=1}^N\mathbb{E}\qty[  \log \frac{\langle e^{\beta\qty(K_{i,1}\sigma_{i,1}\sigma_{i+1,1} + K_{i,2}\sigma_{i,2}\sigma_{i+1,2})}\rangle_{H_{\backslash i}^\mathrm{ladder}(t)} }{ \langle e^{\beta\qty(\hat{l}_{i,1}\sigma_{i,1}\sigma_{i,2}  + \hat{l}_{i,2}\sigma_{i+1,1}\sigma_{i+1,2}  )} \rangle_{H_{\backslash i}^\mathrm{ladder}(t)} }  ]
\no\\
&=&-\frac{1}{\beta }\mathbb{E}\qty[  \log \frac{\langle e^{\beta\qty(K_{i,1}\sigma_{i,1}\sigma_{i+1,1} + K_{i,2}\sigma_{i,2}\sigma_{i+1,2})}\rangle_{H_{\backslash i}^\mathrm{ladder}(t)} }{ \langle e^{\beta\qty(\hat{l}_{i,1}\sigma_{i,1}\sigma_{i,2}  + \hat{l}_{i,2}\sigma_{i+1,1}\sigma_{i+1,2}  )} \rangle_{H_{\backslash i}^\mathrm{ladder}(t)} }  ],
\ee
where we have applied the translation invariance property, and $\langle\cdots \rangle_{H_{\backslash i}^\mathrm{ladder}(t)} $ denotes the thermal average with respect to $H_{\backslash i}^\mathrm{ladder}(t)$.
Here, $H_{\backslash i}^\mathrm{ladder}(t)$ is defined as
\be
&&H_{\backslash i}^\mathrm{ladder}(t)
\no\\
&=& H_{N}^\mathrm{ladder}(t) +M_i(t)(K_{i,1}\sigma_{i,1}\sigma_{i+1,1} + K_{i,2}\sigma_{i,2}\sigma_{i+1,2}) + (1-M_i(t))(\hat{l}_{i,1}\sigma_{i,1}\sigma_{i,2}  + \hat{l}_{i,2}\sigma_{i+1,1}\sigma_{i+1,2}  ). \label{ladder-inter-H-free}
\ee
Note that $H_{\backslash i}^\mathrm{ladder}(t)$ does not contain $K_{i,1}$, $K_{i,2}$, $\hat{l}_{i,1}$, and $\hat{l}_{i,2}$.
The lattice structure of $H_{\backslash i}^\mathrm{ladder}(t)$ is a two-leg ladder with free boundary conditions.

After some calculations based on Eq. (\ref{ladder-cavity-eq}), we obtain
\be
f_N^\mathrm{ladder}(0)+\dv{}{t}f_N^\mathrm{ladder}(t)\big|_{t=0}&=&f_\mathrm{RS}^\mathrm{ladder}. \label{ladder-info-t=0}
\ee
Then, 
\be
f_N^\mathrm{ladder}(1)&=&f_\mathrm{RS}^\mathrm{ladder}+\int_0^1 dt\qty(\dv{}{t}f_N^\mathrm{ladder}(t) -\dv{}{t}f_N^\mathrm{ladder}(t)\big|_{t=0}) .
\ee
Thus, to obtain Theorem 2, it is sufficient to prove 
\be
\lim_{N\to\infty}\int_0^1 dt\qty(\dv{}{t}f_N^\mathrm{ladder}(t) -\dv{}{t}f_N^\mathrm{ladder}(t)\big|_{t=0}) \ge 0.
\ee

We obtain
\be
&&-\beta\qty (\dv{}{t}f_N^\mathrm{ladder}(t) -\dv{}{t}f_N^\mathrm{ladder}(t)\big|_{t=0})
\no\\
&=&\mathbb{E}\qty[  \log   \langle (1+\sigma_{i,1} \sigma_{i+1,1}\tanh\beta K_{i,1})(1+\sigma_{i,2}   \sigma_{i+1,2}    \tanh\beta K_{i,2}  ) \rangle_{H_{\backslash i}^\mathrm{ladder}(t)}  ]
\no\\
&&-\mathbb{E}\qty[   \log  \langle (1+\tanh\beta \hat{l}_{i,1}   \sigma_{i,1}\sigma_{i,2}    ) (1+\tanh\beta \hat{l}_{i,2}   \sigma_{i+1,1}\sigma_{i+1,2}    ) \rangle_{H_{\backslash i}^\mathrm{ladder}(t)} ]
\no\\
&&+\mathbb{E}\qty[   \log   (1+\tanh\beta K_{i,1} \tanh\beta K_{i,2} \tanh\beta ({L}_i+\hat{l}_{i,1})  \tanh\beta ({L}_{i+1}+\hat{l}_{i+1,1})  )  ]  \label{ladder-first-t-0} .
\ee
The proof procedure from this point onwards is identical to that of Theorem 1, albeit with the need for a few additional calculations to expand the logarithm of the three terms in Eq. (\ref{ladder-first-t-0}).
In the thermal average $\langle \cdots \rangle_{H_{\backslash i}^\mathrm{ladder}(t)}$, spins $\sigma_{i,1}\sigma_{i,2}$ and $\sigma_{i+1,1}\sigma_{i+1,2}$ are located at both ends of the two-leg ladder with free boundary conditions, and the probability of each interaction's existence is determined by $t$.
Thus,  the correlation functions are decoupled, 
\be
\langle \sigma_{i,1} \sigma_{i,2} \sigma_{i+1,1} \sigma_{i+1,2} \rangle_{H_{\backslash i}^\mathrm{ladder}(t)}&=&\langle \sigma_{i,1} \sigma_{i,2} \rangle_{H_{\backslash i}^\mathrm{ladder}(t)}\langle \sigma_{i+1,1} \sigma_{i+1,2}\rangle_{H_{\backslash i}^\mathrm{ladder}(t)},
\\
\langle \sigma_{i,1}\sigma_{i+1,1} \rangle_{H_{\backslash i}^\mathrm{ladder}(t)}&=&\langle \sigma_{i,1} \rangle_{H_{\backslash i}^\mathrm{ladder}(t)}\langle \sigma_{i+1,1} \rangle_{H_{\backslash i}^\mathrm{ladder}(t)}=0,
\\
\langle \sigma_{i,2}\sigma_{i+1,2} \rangle_{H_{\backslash i}^\mathrm{ladder}(t)}&=&\langle \sigma_{i,2} \rangle_{H_{\backslash i}^\mathrm{ladder}(t)}\langle \sigma_{i+1,2} \rangle_{H_{\backslash i}^\mathrm{ladder}(t)}=0,
\ee
with probability $1-(1-(1-t)^2)^N$, where we have used $\langle \sigma_{i,1} \rangle_{H_{\backslash i}^\mathrm{ladder}(t)}=0$ owing to spin-flip symmetry.
Then, we have
\be
&&\mathbb{E}\qty[  \log   \langle (1+\sigma_{i,1} \sigma_{i+1,1}\tanh\beta K_{i,1})(1+\sigma_{i,2}   \sigma_{i+1,2}    \tanh\beta K_{i,2}  ) \rangle_{H_{\backslash i}^\mathrm{ladder}(t)}  ]
\no\\
&=&\mathbb{E}\qty[  \log    (1+ \tanh\beta K_{i,1}      \tanh\beta K_{i,2} \langle \sigma_{i,1} \sigma_{i,2}\rangle_{H_{\backslash i}^\mathrm{ladder}(t)} \langle \sigma_{i+1,1}  \sigma_{i+1,2}\rangle_{H_{\backslash i}^\mathrm{ladder}(t)}  )   ]+\mathcal{O}((1-(1-t)^2)^N) , \label{ladder-dilute-evalu-1}
\ee
and
\be
&&-\mathbb{E}\qty[   \log  \langle (1+\tanh\beta \hat{l}_{i,1}   \sigma_{i,1}\sigma_{i,2}    ) (1+\tanh\beta \hat{l}_{i,2}   \sigma_{i+1,1}\sigma_{i+1,2}    ) \rangle_{H_{\backslash i}^\mathrm{ladder}(t)} ]
\no\\
&=&-\mathbb{E}\qty[   \log   (1+\tanh\beta \hat{l}_{i,1}   \langle \sigma_{i,1}\sigma_{i,2}\rangle_{H_{\backslash i}^\mathrm{ladder}(t)}    ) (1+\tanh\beta \hat{l}_{i,2}   \langle \sigma_{i+1,1}\sigma_{i+1,2}  \rangle_{H_{\backslash i}^\mathrm{ladder}(t)}  )  ] +\mathcal{O}((1-(1-t)^2)^N) 
\no\\
&=&-2\mathbb{E}\qty[   \log   (1+ \tanh\beta K_{i,1} \tanh\beta K_{i,2} \tanh\beta (L_i+\hat{l}_{i,1})    \langle \sigma_{i,1}\sigma_{i,2}\rangle_{H_{\backslash i}^\mathrm{ladder}(t)}    )  ] 
+\mathcal{O}((1-(1-t)^2)^N) , \label{ladder-dilute-evalu-2}
\ee
where we have used Eq. (\ref{ladder-cavity-eq}) in the last equality.
By substituting Eqs. (\ref{ladder-dilute-evalu-1}) and (\ref{ladder-dilute-evalu-2}) into Eq. (\ref{ladder-first-t-0}), we obtain
\be
&&-\beta\qty (\dv{}{t}f_N^\mathrm{ladder}(t) -\dv{}{t}f_N^\mathrm{ladder}(t)\big|_{t=0})
\no\\
&=&\mathbb{E}\qty[  \log    (1+ \tanh\beta K_{i,1} \tanh\beta K_{i,2} \langle \sigma_{i,1} \sigma_{i,2}\rangle_{H_{\backslash i}^\mathrm{ladder}(t)} \langle \sigma_{i+1,1}  \sigma_{i+1,2}\rangle_{H_{\backslash i}^\mathrm{ladder}(t)}  )   ]
\no\\
&&-2\mathbb{E}\qty[   \log   (1+ \tanh\beta K_{i,1}  \tanh\beta K_{i,2} \tanh\beta (L_i+\hat{l}_{i,1})    \langle \sigma_{i,1}\sigma_{i,2}\rangle_{H_{\backslash i}^\mathrm{ladder}(t)}    )  ] 
\no\\
&&+\mathbb{E}\qty[   \log   (1+\tanh\beta K_{i,1} \tanh\beta K_{i,2} \tanh\beta ({L}_i+\hat{l}_{i,1})  \tanh\beta ({L}_{i+1}+\hat{l}_{i+1,1})  )  ]   
\no\\
&&+\mathcal{O}((1-(1-t)^2)^N) .
\ee
In addition, given that
\be
|\tanh\beta K_{i,1} \tanh\beta K_{i,2} \langle \sigma_{i,1} \sigma_{i,2}\rangle_{H_{\backslash i}^\mathrm{ladder}(t)} \langle \sigma_{i+1,1}  \sigma_{i+1,2}\rangle_{H_{\backslash i}^\mathrm{ladder}(t)} |<1,
\\
| \tanh\beta K_{i,1}  \tanh\beta K_{i,2} \tanh\beta (L_i+\hat{l}_{i,1})    \langle \sigma_{i,1}\sigma_{i,2}\rangle_{H_{\backslash i}^\mathrm{ladder}(t)} |<1,
\\
|\tanh\beta K_{i,1} \tanh\beta K_{i,2} \tanh\beta ({L}_i+\hat{l}_{i,1})  \tanh\beta ({L}_{i+1}+\hat{l}_{i+1,1})|<1,
\ee
we can expand the logarithm of the three terms as
\be
&&-\beta\qty (\dv{}{t}f_N^\mathrm{ladder}(t) -\dv{}{t}f_N^\mathrm{ladder}(t)\big|_{t=0})
\no\\
&=&-\sum_{n=1}^\infty\frac{1}{2n}\mathbb{E}\qty[\tanh^{2n}\beta K_{i,1} ]\mathbb{E}\qty[\tanh^{2n}\beta K_{i,2} ] \left(\mathbb{E}\qty[   \langle \sigma_{i,1} \sigma_{i,2}\rangle_{H_{\backslash i}^\mathrm{ladder}(t)}^{2n} \langle \sigma_{i+1,1}  \sigma_{i+1,2}\rangle_{H_{\backslash i}^\mathrm{ladder}(t)}^{2n} ] \right.
\no\\
&&\left. -2\mathbb{E}\qty[ \tanh^{2n}\beta ({L}_i+\hat{l}_{i,1})]\mathbb{E}[  \langle \sigma_{i,1} \sigma_{i,2}\rangle_{H_{\backslash i}^\mathrm{ladder}(t)}^{2n}  ] +\mathbb{E}\qty[  \tanh^{2n}\beta ({L}_i+\hat{l}_{i,1})]^2\right)
\no\\
&&+\mathcal{O}((1-(1-t)^2)^N),
\ee
where we have used $\mathbb{E}\qty[\tanh^{2n-1}\beta K_{i,1}  ]=0$ for any positive integer $n$ that follows from $P(K_{i,1}  )=P(-K_{i,1}  )$.
Again, we use the fact that, in the thermal average $\langle \cdots \rangle_{H_{\backslash i}^\mathrm{ladder}(t)}$, both ends of the two-leg ladder are connected with probability $\mathcal{O}((1-(1-t)^2)^N)$, which implies
\be
\mathbb{E}\qty[   \langle \sigma_{i,1} \sigma_{i,2}\rangle_{H_{\backslash i}^\mathrm{ladder}(t)}^{2n} \langle \sigma_{i+1,1}  \sigma_{i+1,2}\rangle_{H_{\backslash i}^\mathrm{ladder}(t)}^{2n} ]
&=&\mathbb{E}\qty[   \langle \sigma_{i,1} \sigma_{i,2}\rangle_{H_{\backslash i}^\mathrm{ladder}(t)}^{2n} ]^2+\mathcal{O}((1-(1-t)^2)^N).
\ee
Then, we obtain
\be
&&-\beta\qty (\dv{}{t}f_N^\mathrm{ladder}(t) -\dv{}{t}f_N^\mathrm{ladder}(t)\big|_{t=0})
\no\\
&=&-\sum_{n=1}^\infty\frac{1}{2n}\mathbb{E}\qty[\tanh^{2n}\beta K_{i,1} ]\mathbb{E}\qty[\tanh^{2n}\beta K_{i,2} ]
\mathbb{E}\qty[   \langle \sigma_{i,1} \sigma_{i,2}\rangle_{H_{\backslash i}^\mathrm{ladder}(t)}^{2n}-  \tanh^{2n}\beta ({L}_i+\hat{l}_{i,1})]^2
\no\\
&&+\mathcal{O}((1-(1-t)^2)^N)
\no\\
&\le&\mathcal{O}((1-(1-t)^2)^N).
\ee
Finally, by taking the thermodynamic limit, we obtain
\be
\lim_{N\to\infty}f_N^\mathrm{ladder}
&=&f_\mathrm{RS}^\mathrm{ladder}+\lim_{N\to\infty}\int_0^1 dt\qty(\dv{}{t}f_N^\mathrm{ladder}(t) -\dv{}{t}f_N^\mathrm{ladder}(t)\big|_{t=0}) 
\no\\
&\ge&f_\mathrm{RS}^\mathrm{ladder}+\lim_{N\to\infty} \int_0^1 dt \mathcal{O}((1-(1-t)^2)^N)
\no\\
&=&f_\mathrm{RS}^\mathrm{ladder},
\ee
which proves Theorem 2.

%%%%%%%%%%%%%%%%%%%%%%%%%%%%%%%%%%%%%%%%%%%%%%%%%%%%%%%%%%%%%%%
\section{Discussions}
Using the interpolation method, we rigorously proved that the RS cavity method provides lower bounds on the free energies of Ising spin glass models in one dimension, such as a one-dimensional chain and a two-leg ladder.
Our proof relies strongly on the property that correlations at both ends are neglected in the thermal average with respect to the one-dimensional interpolating Hamiltonians (\ref{chain-inter-H-free}) and (\ref{ladder-inter-H-free}).
Thus, our method does not extend to more than two dimensions.

The interpolation method was effective in one dimension owing to the validity of Eqs. (\ref{chain-info-t=0}) and (\ref{ladder-info-t=0}).
We emphasize that the left-hand sides of Eqs. (\ref{chain-info-t=0}) and (\ref{ladder-info-t=0}) are equivalent to Oguchi's approximate free energies~\cite{Oguchi} in the case of ferromagnetic interactions.
It is known that Oguchi's approximate free energy coincides with the exact solution of the free energy for ferromagnetic systems in one dimension.
This excellent property is closely related to the equality of Eqs. (\ref{chain-info-t=0}) and (\ref{ladder-info-t=0}).

In one dimension, it is naturally expected that the RS cavity method is rigorous for Ising spin glass models.
Indeed, on a one-dimensional chain, the replicated transfer matrix method allows for proving that the RS cavity formula is rigorous in the thermodynamic limit~\cite{LMR}.
Thus, an important future direction is to prove the converse bound of the present study, especially for a two-leg ladder.

%%%%%%%%%%%%%%%%%%%%%%%%%%%%%%%%%%%%%%%%%%%%%%%%%%%%%%%%%%%%%%%
\section*{Data availability statement}
No new data were created or analysed in this study.

%%%%%%%%%%%%%%%%%%%%%%%%%%%%%%%%%%%%%%%%%%%%%%%%%%%%%%%%%%%%%%%%%%
\section*{Acknowledgements}
This study was supported by JSPS KAKENHI Grant Nos. 24K16973 and 23H01432.
This study received financial support from the Public\verb|\|Private R\&D Investment Strategic Expansion PrograM (PRISM) and programs for Bridging the gap between R\&D and the IDeal society (society 5.0) and Generating Economic and social value (BRIDGE) from Cabinet Office.

%%%%%%%%%%%%%%%%%%%%%%%%%%%%%%%%%%%%%%%%%%%%%%%%%%%%%%%%%%%%%%%%%
\appendix
%%%%%%%%%%%%%%%%%%%%%%%%%%%%%%%%%%%%%%%%%%%%%%%%%%%%%%%%%%%%%%%%%
\section{Derivation of RS cavity formula (\ref{chain-cavity-formula}) on one-dimensional chain} 
Let us consider an Ising spin glass model on a one-dimensional chain with free boundary conditions:
\be
H_{N,\mathrm{free}}^\mathrm{chain}&=&- \sum_{i=1}^{N-1} J_{i} \sigma_i \sigma_{i+1}  - \sum_{i=1}^{N}  H_i\sigma_i .
\ee
First, we assume that, when we take the summation over spin variables from $\sigma_{1}$ to $\sigma_{N-1}$, its contribution is represented by a cavity field $\hat{g}_{N-1}$ and a constant factor $A_{N-1}$ as follows (this assumption that the contribution from $\sigma_{1}$ to $\sigma_{N-1}$ can be expressed in terms of a simple random field corresponds to the RS assumption):
\be
\mathbb{E}\left[\log Z_{N,\mathrm{free}}^\mathrm{chain} \right]&=&\mathbb{E}\left[A_{N-1} \log  \Tr_{\sigma_N}\qty( e^{\beta (\hat{g}_{N-1}+H_N)\sigma_N})\right]
\no\\
&=&\mathbb{E}\left[A_{N-1} \log  2\cosh\beta (\hat{g}_{N-1}+H_N) \right],
\ee
where $\Tr_{\sigma_N}$ is the summation over $\sigma_N$.
By repeating the same assumption for a system with one more spin number, we obtain
\be
\mathbb{E}\left[\log Z_{N+1,\mathrm{free}}^\mathrm{chain} \right]&=&\mathbb{E}\left[A_{N-1} \log  \Tr_{\sigma_N,\sigma_{N+1}} \qty(e^{\beta (\hat{g}_{N-1}+H_N)\sigma_N+ \beta J_N\sigma_N\sigma_{N+1} +\beta H_{N+1}\sigma_{N+1} })\right], 
\ee
where $\Tr_{\sigma_N, \sigma_{N+1}}$ is the summation over $\sigma_N$ and $\sigma_{N+1}$.
Let us introduce the new cavity field $\hat{g}_N$, which satisfies
\be
\tanh\beta \hat{g}_N &\stackrel{d}{=}&\tanh\beta J_N\tanh\beta (\hat{g}_{N-1}+H_N),
\ee
where the equality holds in distribution.
Then, we can rewrite $\mathbb{E}\left[\log Z_{N+1,\mathrm{free}}^\mathrm{chain} \right]$ as
\be
\mathbb{E}\left[\log Z_{N+1,\mathrm{free}}^\mathrm{chain} \right]&=&\mathbb{E}\left[A_{N-1}  \log \frac{ 4 \cosh\beta(\hat{g}_{N-1}+H_{N}) \cosh\beta J_N \cosh\beta (\hat{g}_{N}+H_{N+1}) }{\cosh\beta\hat{g}_N}\right].
\ee
Thus, we have
\be
\mathbb{E}\left[\log Z_{N+1,\mathrm{free}}^\mathrm{chain} \right]-\mathbb{E}\left[\log Z_{N,\mathrm{free}}^\mathrm{chain} \right]
&=&\mathbb{E}\left[  \log \frac{ 2 \cosh\beta J_N \cosh\beta (\hat{g}_{N}+H_{N+1}) }{\cosh\beta\hat{g}_N}\right].
\ee
By taking the thermodynamic limit $N\to\infty$, the probability distributions of $\hat{g}_{N-1}$ and $\hat{g}_{N}$ are expected to converge to the same one, implying
\be
\lim_{N\to\infty}\qty(\mathbb{E}\left[\log Z_{N+1,\mathrm{free}}^\mathrm{chain} \right]-\mathbb{E}\left[\log Z_{N,\mathrm{free}}^\mathrm{chain} \right])
&=&\mathbb{E}\left[  \log \frac{ 2 \cosh\beta J \cosh\beta (\hat{g}_{}+H_{}) }{\cosh\beta\hat{g}}\right], \label{appeA-RS-formula}
\ee
with
\be
\tanh\beta \hat{g} &\stackrel{d}{=}&\tanh\beta J \tanh\beta (\hat{g}_{}'+H), \label{appeA-RS-formula2}
\ee
where the equality holds in distribution, $J$ is a random variable following the same distribution as $J_i$, $H$ is a random variable following the same distribution as $H_i$, and $\hat{g}'$ is a random variable following the same distribution as $\hat{g}$.

Finally, because $\mathbb{E}\left[\log Z_{N+1,\mathrm{free}}^\mathrm{chain} \right]-\mathbb{E}\left[\log Z_{N,\mathrm{free}}^\mathrm{chain} \right]$ is considered the free energy density per spin, we assume the following relation (see Ref. \cite{ASS} for a mathematically rigorous treatment):
\be
\lim_{N\to\infty} \frac{1}{N}\mathbb{E}\left[\log Z_{N,\mathrm{free}}^\mathrm{chain} \right]
&=&\lim_{N\to\infty}\qty(\mathbb{E}\left[\log Z_{N+1,\mathrm{free}}^\mathrm{chain} \right]-\mathbb{E}\left[\log Z_{N,\mathrm{free}}^\mathrm{chain} \right]).
\ee
In combination with Eqs. (\ref{appeA-RS-formula}) and (\ref{appeA-RS-formula2}), we obtain the RS cavity formula (\ref{chain-cavity-formula}).

%%%%%%%%%%%%%%%%%%%%%%%%%%%%%%%%%%%%%%%%%%%%%%%%%%%%%%%%%%%%%%%%%
%%%%%%%%%%%%%%%%%%%%%%%%%%%%%%%%%%%%%%%%%%%%%%%%%%%%%%%%%%%%%%%%%
\section{Derivation of RS cavity formula (\ref{ladder-cavity-formula}) on two-leg ladder} 
Let us consider an Ising spin glass model on a two-leg ladder with free boundary conditions:
\be
H_{N,\mathrm{free}}^\mathrm{ladder}&=&-\sum_{i=1}^{N} L_{i}\sigma_{i,1}\sigma_{i,2} - \sum_{i=1}^{N-1}(K_{i,1}\sigma_{i,1}\sigma_{i+1,1} + K_{i,2}\sigma_{i,2}\sigma_{i+1,2}).
\ee
First, we assume that, when we take the summation over spin variables from $\sigma_{1,1}$ and $\sigma_{1,2}$ to $\sigma_{N-1,1}$ and $\sigma_{N-1,2}$, its contribution is represented by a cavity coupling $\hat{s}_{N-1}$ and a constant factor $B_{N-1}$ as follows:
\be
\mathbb{E}\left[\log Z_{N,\mathrm{free}}^\mathrm{ladder} \right]&=&\mathbb{E}\left[B_{N-1} \log  \Tr_{\sigma_N}\qty( e^{\beta (\hat{s}_{N-1}+L_N)\sigma_{N,1}\sigma_{N,2}})\right]
\no\\
&=&\mathbb{E}\left[B_{N-1} \log  4\cosh\beta (\hat{s}_{N-1}+L_N) \right],
\ee
where $\Tr_{\sigma_N}$ is the summation over $\sigma_{N,1}$ and $\sigma_{N,2}$.
By repeating the same assumption for a system with one more spin number, we obtain
\be
\mathbb{E}\left[\log Z_{N+1,\mathrm{free}}^\mathrm{ladder} \right]&=&\mathbb{E}\left[B_{N-1} \log  \Tr_{\sigma_N,\sigma_{N+1}}\qty( e^{\beta (\hat{s}_{N-1}+L_N)\sigma_{N,1}\sigma_{N,2}+\beta(K_{N,1}\sigma_{N,1}\sigma_{N+1,1} + K_{N,2}\sigma_{N,2}\sigma_{N+1,2}) +\beta L_{N+1}\sigma_{N+1,1}\sigma_{N+1,2}})\right], 
\no\\
\ee
where $\Tr_{\sigma_N, \sigma_{N+1}}$ is the summation over $\sigma_{N,1}$, $\sigma_{N,2}$, $\sigma_{N+1,1}$, and $\sigma_{N+1,2}$.
Let us introduce the new cavity coupling $\hat{s}_N$, which satisfies
\be
\tanh\beta \hat{s}_N  &\stackrel{d}{=}& \tanh\beta K_{N,1} \tanh\beta K_{N,2} \tanh\beta (\hat{s}_{N-1}+L_N) , 
\ee
where the equality holds in distribution.
Then, we can rewrite $\mathbb{E}\left[\log Z_{N+1,\mathrm{free}}^\mathrm{ladder} \right]$ as
\be
\mathbb{E}\left[\log Z_{N+1,\mathrm{free}}^\mathrm{ladder} \right]&=&\mathbb{E}\left[B_{N-1}  \log \frac{ 16 \cosh\beta(\hat{s}_{N-1}+L_{N}) \cosh\beta K_{N,1}\cosh\beta K_{N,2} \cosh\beta (\hat{s}_{N}+L_{N+1}) }{\cosh\beta\hat{s}_N}\right].
\ee
Thus, we have
\be
\mathbb{E}\left[\log Z_{N+1,\mathrm{free}}^\mathrm{ladder} \right]-\mathbb{E}\left[\log Z_{N,\mathrm{free}}^\mathrm{ladder} \right]
&=&\mathbb{E}\left[  \log \frac{ 4  \cosh\beta K_{N,1}\cosh\beta K_{N,2} \cosh\beta (\hat{s}_{N}+L_{N+1}) }{\cosh\beta\hat{s}_N}\right].
\ee
By taking the thermodynamic limit $N\to\infty$, the probability distributions of $\hat{s}_{N-1}$ and $\hat{s}_{N}$ are expected to converge to the same one.
This implies
\be
\lim_{N\to\infty}\qty(\mathbb{E}\left[\log Z_{N+1,\mathrm{free}}^\mathrm{ladder} \right]-\mathbb{E}\left[\log Z_{N,\mathrm{free}}^\mathrm{ladder} \right])
&=&\mathbb{E}\left[  \log \frac{ 4  \cosh\beta K_1\cosh\beta K_2 \cosh\beta (\hat{s}+L_{}) }{\cosh\beta\hat{s}}\right],
\ee
with
\be
\tanh\beta \hat{s}  &\stackrel{d}{=}& \tanh\beta K_1 \tanh\beta K_2 \tanh\beta (\hat{s}'+L) ,
\ee
where the equality holds in distribution, $K_1$ is a random variable following the same distribution as $K_{i,1}$, $K_2$ is a random variable following the same distribution as $K_{i,2}$, $L$ is a random variable following the same distribution as $L_i$, and $\hat{s}'$ is a random variable following the same distribution as $\hat{s}$.

By contrast, given that $\mathbb{E}\left[\log Z_{N+1,\mathrm{free}}^\mathrm{ladder} \right]-\mathbb{E}\left[\log Z_{N,\mathrm{free}}^\mathrm{ladder} \right]$ is considered the free energy density per spin, we assume the following relation:
\be
\lim_{N\to\infty} \frac{1}{N}\mathbb{E}\left[\log Z_{N,\mathrm{free}}^\mathrm{ladder} \right]
&=&\lim_{N\to\infty}\qty(\mathbb{E}\left[\log Z_{N+1,\mathrm{free}}^\mathrm{ladder} \right]-\mathbb{E}\left[\log Z_{N,\mathrm{free}}^\mathrm{ladder} \right]).
\ee
Therefore, we obtain the RS cavity formula (\ref{ladder-cavity-formula}).

%%%%%%%%%%%%%%%%%%%%%%%%%%%%%%%%%%%%%%%%%%%%%%%%%%%%%%%%%%%%%
%%%%%%%%%%%%%%%%%%%%%%%%%%%%%%%%%%%%%%%%%%%%%%%%%%%%%%%%%%%%%%%%%%%%%%%%%%%%


\begin{thebibliography}{99}



\bibitem{Nishimori}
H. Nishimori,
\textit{Statistical Physics of Spin Glasses and Information Processing: An Introduction}
 (Oxford University Press, Oxford, 2001).

\bibitem{MM}
A. Montanari and M. M\'{e}zard, 
\textit{Information, Physics and Computation} (Oxford Univ. Press, 2009).

\bibitem{Parisi}
G. Parisi,
{A sequence of approximated solutions to the S-K model for spin glasses},
J. Phys. A: Math. Gen. \textbf{13}, L115 (1980).


\bibitem{GT}
F. Guerra and F. L. Toninelli,
{The Thermodynamic Limit in Mean Field Spin Glass Models},
Commun. Math. Phys. \textbf{230}, 71 (2002).



\bibitem{Guerra}
F. Guerra,
{Broken Replica Symmetry Bounds in the Mean Field Spin Glass Model},
Commun. Math. Phys. \textbf{233}, 1 (2003).

\bibitem{Talagrand}
M. Talagrand,
{The Parisi formula},
Ann. Math. \textbf{163}, 221 (2006).

\bibitem{Panchenko}
D. Panchenko,
{The Parisi formula for mixed $p$-spin models},
Ann. Probab. \textbf{42}, 946 (2014).


\bibitem{MG}
M. M\'{e}zard and G. Parisi,
{The Bethe lattice spin glass revisited},
Euro. Phys. J. B, \textbf{233}, 217 (2001).

%%%%%%%%%%%%%%%%%%%%%%%%%%%%%%%%%%%%%%%%%%%%%
%%希釈平均場のバウンド
\bibitem{FL}
S. Franz and M. Leone,
{Replica bounds for optimization problems and diluted spin systems},
J. Stat. Phys. \textbf{111}, 535 (2003).

\bibitem{FLT}
S. Franz and M. Leone, and F. L. Toninelli,
{Replica bounds for diluted non-Poissonian spin systems},
J. Phys. A: Math. Gen. \textbf{36}, 10967 (2003).

\bibitem{PT}
D. Panchenko and M. Talagrand,
{Bounds for diluted mean-fields spin glass models},
Probab. Theory Relat. Fields \textbf{130}, 319 (2004).





%%%%%%%%%%%%%%%%%%%%%%%%%%%%%%%%%%%%%%%%%%%%%_
%%希釈平均場の進展
\bibitem{GT2}
F. Guerra and F. L. Toninelli,
{The High Temperature Region of the Viana-Bray Diluted Spin Glass Model},
J. Stat. Phys. \textbf{115}, 531 (2004).

\bibitem{Panchenko2}
D. Panchenko,
{Spin glass models from the point of view of spin distributions},
Ann. Probab. \textbf{41}, 1315 (2013).

\bibitem{Panchenko3}
D. Panchenko,
{Structure of 1-RSB asymptotic Gibbs measures in the diluted p-spin models},
J. Stat. Phys. \textbf{155}, 1 (2014).

\bibitem{Panchenko4}
D. Panchenko,
{Hierarchical exchangeability of pure states in mean field spin glass models},
Probab. Theory Relat. Fields \textbf{161}, 619 (2015).


\bibitem{DSS}
J. Ding, A. Sly, and N. Sun,
{Maximum independent sets on random regular graphs}, 
Acta Math. 217, 263 (2016).
%d->無限大で1RSBが厳密であることの証明

\bibitem{LO}
M. Lelarge and M. Oulamara,
{Replica bounds by combinatorial interpolation for diluted spin systems}, 
J. Stat. Phys. 173, 917 (2018).
%ランダム正則グラフでのレプリカバウンド

\bibitem{OP}
A. Coja-Oghlan and W. Perkins,
{Spin systems on Bethe lattices},
Commun. Math. Phys. \textbf{372}, 441 (2019).

\bibitem{Harangi}
V. Harangi,
{Improved Replica Bounds for the Independence Ratio of Random Regular Graphs},
J. Stat. Phys. 190, 60 (2023).



%%%%%%%%%%%%%%%%%%%%%%%%%%%%%%%%%%%%%%%%%%%%%
%一次元スピングラスの転送行列法
\bibitem{DVP}
B. Derrida, J. Vannimenus, and Y. Pomeau,
{Simple frustrated systems: chains, strips and squares},
J. Phys. C \textbf{11}, 4749 (1978).

\bibitem{LFN}
J. M. Luck, M. Funke, and T. M. Nieuwenhuizen, 
{Low-temperature thermodynamics of random-field Ising chains: exact results},
J. Phys. A \textbf{24}, 4155 (1991).

\bibitem{WM}
M. Weigt and R. Monasson,
{Replica structure of one-dimensional disordered Ising models},
Europhys. Lett. \textbf{36}, 209 (1996).


\bibitem{BM}
A. J. Bray and M. A. Moore,
{Finite size effects in spin glass overlap functions},
J. Phys. A: Math. Gen. \textbf{18}, L683 (1985).


\bibitem{LMR}
C. Lucibello, F. Morone, and T. Rizzo,
{One-dimensional disordered Ising models by replica and cavity methods},
Phys. Rev. E \textbf{90}, 012140 (2014).


\bibitem{Oguchi}
A. Oguchi,
{Approximate Method for the Free Energy},
Prog.  Theor. Phys. \textbf{56}, 1442 (1976).



%%%%%%%%%%%%%%%%%%%%%%%%%
%短距離相互作用の熱力学極限の存在証明
\bibitem{Vuillermat}
P. A. Vuillermat, 
{Thermodynamics of quenched random spin systems, and application to the problem of phase transitions in magnetic (spin) glasses},
J. Phys. A \textbf{10}, 1319 (1977).

\bibitem{PF}
L. A. Pastur and A. L. Figotin,
{Theory of disordered spin systems},
Teor. Mat. Fiz. \textbf{35}, 193 (1978).






\bibitem{Matsubara}
F. Matsubara,
{Theory of One-Dimensional Random Mixture of Ising Spins},
Prog.  Theor. Phys. \textbf{51}, 1694 (1974).
%一次元スピングラスの有効場近似

\bibitem{KF}
S. Katsura and S. Fujiki,
{Distribution of spins and the thermodynamic properties in the glass-like (spin glass) phase of random Ising bond models},
J. Phys. C: Solid State Phys. \textbf{12,} 1087 (1979).
%スピングラスの有効場近似、一般的な設定でのBethe近似



\bibitem{ASS}
M. Aizenman, R. Sims, and S. L. Starr,
{Extended variational principle for the Sherrington-Kirkpatrick spin-glass model},
Phys. Rev. B \textbf{68}, 214403 (2003).



\end{thebibliography}
\end{document}